\documentclass[letterpaper,journal]{IEEEtran}

\usepackage{amsmath}
\usepackage{graphicx}
\usepackage[nolist]{acronym}
\usepackage[hidelinks]{hyperref}
\usepackage{float}

\usepackage{multirow}
\usepackage{booktabs}
\usepackage[table]{xcolor}
\usepackage[skip=0pt]{caption}

\usepackage[markup=default]{changes}
\definechangesauthor[name={R1}, color=blue]{R1}
\definechangesauthor[name={R2}, color=red]{R2}
\definechangesauthor[name={R3}, color=orange]{R3}

\newacro{3GPP}{Third Generation Partnership Project}
\newacro{5G}{5th Generation}
\newacro{6G}{6th Generation}
\newacro{AoA}{Angle of Arrival}
\newacro{AoD}{Angle of Departure}
\newacro{AR}{Auto-regressive}
\newacro{B5G}{Beyond 5G}
\acrodefplural{BBU}[BBUs]{Baseband Units}
\newacro{BBU}{Baseband Unit}
\newacro{BS}{Base Station}
\newacro{BT}{Beam-Tracking}
\newacro{CNN}{Convolutional Neural Network}
\newacro{CSI}{Channel State Information}
\newacro{CSI-RS}{Channel State Information Reference Signals}
\newacro{CSV}{Comma-Separated Values}
\newacro{DL}{Downlink}
\newacro{GNSS}{Global Navigation Satellite System}
\newacro{HDF5}{Hierarchical Data Format - Version 5}
\newacro{IBC}{Image-Based Coding}
\newacro{I2I}{Infrastructure-to-Infrastructure}
\newacro{L1-RSRP}{Layer 1 Reference Signal Received Power}
\newacro{LOS}{Line-of-Sight}
\newacro{LMS}{Least Mean Squares}
\newacro{LSTM}{Long Short-Term Memory}
\newacro{MAFD}{Mean Absolute First Difference}
\newacro{MIMO}{Multiple-Input Multiple-Output}
\newacro{ML}{Machine Learning}
\newacro{mmWave}{Millimeter Wave}
\newacro{MPC}{Multi-path Components}
\newacro{MSE}{Mean Squared Error}
\newacro{NLMS}{Normalized Least Mean Squares}
\newacro{NLOS}{Non-Line-of-Sight}
\newacro{NN}{Neural Network}
\newacro{NMSE}{Normalized Mean Square Error}
\newacro{OSM}{OpenStreetMap}
\newacro{PHY}{Physical}
\newacro{RAN}{Radio Access Network}
\newacro{RAN-1}{Radio Access Network - 1}
\newacro{RBVC}{Radio Bit-Vector Compression}
\newacro{RCC}{Radio Resource Control}
\acrodefplural{RCC}[RCC]{Radio Resource Control}
\newacro{RF}{Radio Frequency}
\newacro{RIS}{Reconfigurable Intelligent Surface}
\newacro{RNN}{Recurrent Neural Network}
\newacro{RSRP}{Reference Signal Received Power}
\newacro{SINR}{Signal to Interference \& Noise Ratio}
\newacro{RRU}{Remote Radio Unit}
\acrodefplural{RRU}[RRUs]{Remote Radio Units}
\newacro{RT}{Ray-Tracing}
\newacro{Rx}{Receiver}
\newacro{SNR}{Signal-to-Noise Ratio}
\newacro{SUMO}{Simulator for Urban Mobility}
\newacro{THz}{Terahertz}
\newacro{TR}{Throughput Ratio}
\newacro{Tx}{Transmitter}
\newacro{UE}{User Equipment}
\acrodefplural{UE}[UEs]{User Equipments}
\newacro{UCA}{Uniform Circular Array}
\newacro{ULA}{Uniform Linear Array}
\newacro{UPA}{Uniform Planar Array}
\newacro{V2I}{Vehicle-to-Infrastructure}
\newacro{V2V}{Vehicle-to-Vehicle}
\newacro{WI}{Wireless Insite}
\newacro{WLAN}{Wireless Local Area Network}
\newacro{GRU}{Gated Recurrent Unit}
\newacro{SSB}{Synchronization Signal Block}
\newacro{DFT}{Discrete Fourier Transform}
\newacro{MOR}{Measurement Overhead Reduction}

\usepackage[table]{xcolor}
\usepackage{array}
\usepackage[dvipsnames]{xcolor}
\usepackage{amssymb}

\newcolumntype{L}[1]{>{\raggedright\let\newline\\\arraybackslash\hspace{0pt}}m{#1}}
\newcolumntype{C}[1]{>{\centering\let\newline\\\arraybackslash\hspace{0pt}}m{#1}}
\newcolumntype{R}[1]{>{\raggedleft\let\newline\\\arraybackslash\hspace{0pt}}m{#1}}

\begin{document}

\title{DL-Based Beam Management for mmWave Vehicular Networks Exploring Temporal Correlation}

\author{\IEEEauthorblockN{
    Ailton Oliveira\IEEEauthorrefmark{1},
    Amir Khatibi\IEEEauthorrefmark{2},
    Daniel Suzuki\IEEEauthorrefmark{1}, 
    Ilan Correa\IEEEauthorrefmark{1}, 
    José Rezende\IEEEauthorrefmark{2} and
    Aldebaro Klautau\IEEEauthorrefmark{1}} \\
    \IEEEauthorblockA{
        \IEEEauthorrefmark{1} LASSE - Telecom., Automation and Electronics Research and Development Center, Belém-PA, Brazil   \\
        \IEEEauthorrefmark{2} Universidade Federal do Rio de Janeiro, Rio de Janeiro-RJ, Brazil \\
        Email: \IEEEauthorrefmark{1}\{ailton.pinto, daniel.suzuki\}@itec.ufpa.br, ilan@ufpa.br, aldebaro@ufpa.br \\
        \IEEEauthorrefmark{2}\{amir.khatibi, rezende\}@land.ufrj.br}

}

\markboth{Journal of \LaTeX\ Class Files,~Vol.~14, No.~8, August~2015}%
{Shell \MakeLowercase{\textit{et al.}}: Bare Demo of IEEEtran.cls for IEEE Journals}

\maketitle

\begin{abstract}
Millimeter wave communications are essential for modern wireless networks. It supports high data rates but suffers from severe path loss, which requires precise beam alignment to maintain reliable links. This beam management is particularly challenging in highly dynamic scenarios such as vehicle-to-infrastructure, and several methods have been presented. In this work, we propose a deep learning-based beam tracking framework based on sequential prediction using recurrent neural networks with an autoregressive inference strategy that reduces measurement overhead. The proposed architecture can support deep learning models trained for both classification and regression. In contrast to many existing studies that evaluate beam tracking under predominantly line-of-sight (LOS) conditions, our work explicitly includes highly challenging non-LOS scenarios - with up to 50\% non-LOS  incidence in certain datasets - to rigorously assess model robustness. Experimental results demonstrate that our approach maintains high top-\textit{K} accuracy, even under adverse conditions, while reducing the beam measurement overhead by up to 66\%. 
\end{abstract}


\IEEEpeerreviewmaketitle



\section{Introduction}
\label{sec:introduction}


One critical enabler of 6G is the efficient use of \ac{mmWave} (e.g., 28 GHz, 60 GHz) and \ac{THz} frequency bands, which offer abundant spectral resources that can facilitate ultra-high throughput and low latency~\cite{Roh14}. However, these bands suffer from higher path loss and limited diffraction capability, making signal degradation a significant concern in practical deployments. As a result, ensuring reliable connectivity in these bands demands innovations in beam management.
To address these challenges, \ac{MIMO} technologies—particularly massive MIMO—have become foundational in modern wireless systems~\cite{mmWave2020}. Beamforming, a key component of massive \ac{MIMO}, concentrates signal energy along preferred spatial directions, significantly improving link budget and mitigating high-frequency attenuation~\cite{Bjornson19}. The effectiveness of beamforming, however, depends on maintaining precise alignment between the transmitter and receiver beams. This is especially complex in mobile and dynamic scenarios such as \ac{V2I} communications, where misalignment can lead to severe link degradation or complete outage~\cite{yi2024beam}.

Recent \ac{3GPP} efforts~\cite{3gpp2306199} have emphasized the importance of intelligent beam management, which includes not only initial alignment, but also real-time tracking and recovery from failures. Machine learning techniques—especially deep learning—have emerged as promising solutions, offering the potential to learn temporal and spatial signal patterns to guide beam selection and tracking more efficiently than traditional optimization algorithms~\cite{xue2024survey}.

Continuous beam \emph{tracking} and enhanced beam management can yield significant system-level gains, such as reduced energy consumption, faster reconnection times, and lower signaling overhead. For example, smart tracking may reduce beam measurement overhead by
approximately 66\% compared to exhaustive beam sweeping in certain configurations (Section~\ref{sec:experimental_results}), with strategies
ranging from 50\% to 75\% depending on the prediction-to-measurement ratio.

Motivated by these potential gains, this work explores a novel deep learning-based architecture for real-time beam tracking in \ac{mmWave} \ac{MIMO} systems, with focus on \ac{V2I} scenarios. Our approach balances computational efficiency and prediction accuracy, leveraging \ac{RNN}-based models to exploit sequential signal patterns. A measurement substitution strategy based on autoregressive inference further reduces beam measurement overhead without significant accuracy degradation. The contributions of this paper are:
\begin{itemize}

    \item The design and evaluation of two complementary models, an \textbf{RNN Index Classification} and an \textbf{RNN RSRP Regression}, which respectively formulate beam tracking as classification and regression tasks.
    
    \item A comprehensive \textbf{evaluation framework} across realistic vehicular datasets at real-world scenarios, highlighting performance under varying \ac{LOS}/\ac{NLOS} conditions, and introducing a new metric—\textbf{\ac{MAFD}}—to characterize beam dynamics in these scenarios.
    
    \item An analysis of \textbf{measurement substitution strategies}, demonstrating how replacing beam measurements with predictions can reduce sensing overhead by 50\% to 75\%, depending on the replacement ratio, while incurring only marginal loss in accuracy.
\end{itemize}

It is worth noting that position-aware spatial beam pre-selection was explored in preliminary experiments, showing promise in reducing search space with minimal accuracy loss. However, a systematic evaluation of this component is deferred to future work; the results reported in this paper do not rely on any pre-selection stage.

The study is guided by the following research questions:


    \textbf{RQ1:} Considering classification and regression formulations, which supports better generalization and robustness under dynamic channel conditions?

    How does the replacement of real measurements with predicted values affect tracking accuracy, and what are the trade-offs between measurement overhead reduction and prediction performance?

    \textbf{RQ3:} Can the proposed architecture maintain high performance in challenging \ac{NLOS} scenarios, which are frequent in urban \ac{V2I} environments?


The remainder of this paper is organized as follows. Section~\ref{sec:related_work} reviews related work on beam tracking and deep learning for wireless systems. Problem statement is described in Section~\ref{problem_statement}. Section~\ref{sec:dataset_description} characterizes the datasets used. Then in Section~\ref{sec:proposed_architecture}, we detail the methodology, our proposed model architecture and pre-processing strategy. Section~\ref{sec:experimental_results} presents performance evaluations and analyses. Finally, Section~\ref{sec:conclusion} concludes the paper and outlines directions for future research.

\section{Related Work}
\label{sec:related_work}

Traditional beam management relies on heuristic-based strategies standardized by the \ac{3GPP}, which include the brute-force \emph{beam sweeping}~\cite{beam_Sweeping_3gpp.38.802}. These procedures enable the identification of optimal beams through periodic transmission and evaluation of \ac{SSB} and \ac{CSI-RS}~\cite{beam_management_csi_3gpp.38.213}.
Such exhaustive or grid-based searches often lead to significant signaling overhead and latency—particularly in high-mobility settings.
In response, a growing body of research has investigated algorithmic optimizations and learning-based alternatives aimed at accelerating the beam tracking process while preserving, or even enhancing, alignment accuracy.

Beam management has been extensively studied under different methodological paradigms, ranging from standardized heuristic-based strategies to analytical models and machine learning approaches. These methods differ in terms of their reliance on predefined codebooks, adaptability to mobility patterns, computational cost, and data requirements. 


To frame our contribution, this section is organized into three parts: first, \textit{Traditional Methods} standardized by the \ac{3GPP} (Section~\ref{subsec:traditional}); second, \textit{Analytical Methods} that apply model-based prediction techniques (Section~\ref{subsec:analytical}); and finally, \textit{Learning-Based Approaches} that leverage data-driven models for beam tracking (Section~\ref{subsec:learning_based}). 

Table~\ref{tab:summary_related_works} provides a consolidated view of key representative works, comparing their algorithmic category, underlying technique, data sources, and how our method advances the state of the art. This organization facilitates direct comparison between different paradigms, highlighting trade-offs in terms of computational efficiency, robustness to \ac{NLOS} conditions, and reliance on specific types of input data. Unlike many prior works that require long temporal sequences or site-specific sensing information, our approach: a) employs compact temporal sequences (as short as four time steps), b) relies only on historical beam data and estimated \ac{UE} position—readily available in standard \ac{3GPP} architectures, and c) evaluates performance in scenarios with high \ac{NLOS} incidence, demonstrating robustness in challenging environments.

\subsection{Traditional Methods}
\label{subsec:traditional}

Traditional beam management methods, as specified in the \ac{3GPP} standards (e.g., TS 38.213, TS 38.214, TS 38.215), form the foundation of initial access and beam tracking in modern wireless systems. Two key procedures underpin these methods: \textbf{beam sweeping} and \textbf{beam measurement}.

\textbf{Beam sweeping} involves transmitting and/or receiving predefined beams sequentially across a wide angular space, typically using a codebook-based beamforming strategy~\cite{beam_Sweeping_3gpp.38.802}. In downlink, for example, the \ac{BS} periodically broadcasts \ac{SSB}s in different spatial directions, enabling the \ac{UE} to detect candidate beams.

Once candidate beams are detected, \textbf{beam measurement} is performed by the \ac{UE} based on received \ac{SSB}s or \ac{CSI-RS}~\cite{beam_management_csi_3gpp.38.213}, using metrics such as \ac{RSRP} and \ac{SINR} to quantify link quality. These measurements are reported back to the network or used locally for uplink decisions.

Although these procedures are highly structured and reliable, their performance is bounded by the overhead associated with frequent measurements and reporting. These limitations have motivated the exploration of data-driven and learning-based alternatives that can anticipate beam transitions and reduce reliance on exhaustive scanning.


\subsection{Analytical Methods}
\label{subsec:analytical}

Beyond standardized procedures, a variety of analytical methods have been proposed to enhance beam tracking by leveraging recursive estimators and adaptive algorithms. These techniques aim to predict the optimal beam direction based on state evolution models, typically incorporating position, velocity, and signal observations. Unlike learning-based methods, they operate without the need for large datasets or training phases, making them attractive for real-time applications with constrained computational budgets.

Shaham et al.~\cite{shaham2020extended} proposed an Extended Kalman Filter using position, velocity, and channel coefficients as state variables. While computationally efficient, this approach relies on a constant velocity assumption, limiting its applicability in vehicular environments with complex mobility patterns.

Asi et al.~\cite{asi2021beam} compared Least Mean Squares (LMS) and Normalized LMS (NLMS) algorithms. While the NLMS model achieved faster convergence, the lack of systematic step-size selection constrained its adaptability. Similarly, Yi et al.~\cite{yi2024beam} proposed a recursive beam refinement strategy using multi-resolution codebooks to reduce search complexity by exploiting spatial and temporal dependencies. These methods are computationally lightweight but typically depend on well-tuned hyperparameters or simplifying assumptions.

\subsection{Learning-Based Approaches}
\label{subsec:learning_based}

A comprehensive survey by Xue et
al.~\cite{xue2024survey} categorizes AI-based beam management into independent and collaborative training paradigms, covering supervised learning, reinforcement learning, federated learning, and transfer learning approaches. Their analysis identifies two key open challenges that motivate the present work: (i)~the need for compact models that generalize across diverse propagation environments without requiring site-specific retraining, and (ii)~the lack of evaluation under challenging \ac{NLOS} conditions in most existing studies.

Zhao et al.~\cite{zhao2024lstm} used a standard \ac{LSTM} model trained with sub-6 GHz CSI to guide \ac{mmWave} beam selection. Although effective, their model requires long input sequences (16 time steps) and struggles with generalization across heterogeneous \ac{BS} deployments. Similarly, Lim et al.~\cite{lim2021deep} proposed an \ac{LSTM} combined with a Bayesian filter to estimate AoA/AoD variations. While accurate, this method is computationally intensive due to sequential filtering.

Alwakeel et al.~\cite{alwakeel20256g} proposed a machine learning-based framework for beamforming virtualization in 6G systems using software-defined networking and reinforcement learning. Their solution focuses on improving beam management efficiency and energy savings in centralized \ac{LOS} scenarios with limited mobility. While their approach is well suited for infrastructure-level virtualization, our work addresses real-time beam tracking under both \ac{LOS} and \ac{NLOS} conditions in dynamic vehicular networks.

Several recent studies have explored vision- or sensor-aided learning. Zhong et al.~\cite{zhong2024image} and Suzuki et al.~\cite{suzuki2022ray} incorporated CNNs for LIDAR-based prediction, later enhanced by Oliveira et al.~\cite{oliveira2024tracking} to add multimodal data fusion. While accurate in specific environments, these models are often site-specific and require large amounts of heterogeneous data, raising concerns about generalization and training overhead.

Mollah et al.~\cite{mollah2026transformer} proposed a Transformer-based multi-modal fusion framework for \ac{mmWave} beam selection in vehicular scenarios, employing multi-head cross-modal attention across GPS, camera, and LiDAR inputs to predict top-$k$ beams. While their approach achieves strong prediction accuracy and reduces beam searching overhead by over 76\%, it requires multiple synchronized sensing modalities and high-dimensional inputs. In contrast, our method relies solely on beam-domain \ac{RSRP} history, making it significantly lighter and more suitable for deployment in resource-constrained settings.

Recent work by Jiang et al.~\cite{Alkhateeb_baseline} proposed a beam tracking system evaluated on the DeepSense 6G dataset. Their approach, based on GRU networks, achieved up to 95.6\% Top-5 accuracy. However, their evaluation is limited to \ac{LOS} scenarios, which may not fully represent the challenges encountered in more complex or obstructed environments. In this work, we reproduce their techniques on our datasets under assumptions similar to the ones adopted in~\cite{Alkhateeb_baseline}, including the one of perfect knowledge of previously selected optimal beams.

\begin{table*}[h]
    \centering
    \footnotesize
    \caption{Summary of related work and comparison with our contributions}
    \label{tab:summary_related_works}
    \begin{tabular}{L{2.2cm} C{1.2cm} L{3.5cm} L{3cm} L{6cm}}
        \hline
        \textbf{Related Work} & \textbf{Machine Learning} & \textbf{Technique} & \textbf{Data Source} & \textbf{Our Contribution Relative to Work} \\
        \hline
        Yi et al.~\cite{yi2024beam} & $\times$ & Recursive multi-resolution search & Hierarchical codebooks & Data-driven beam filtering strategy, not reliant on fixed heuristics \\
        Shaham et al.~\cite{shaham2020extended} & $\times$ & Extended Kalman Filter & Kinematic model (position/velocity) & Handles non-constant velocity with more realistic vehicular dynamics \\
        Asi et al.~\cite{asi2021beam} & $\times$ & LMS vs. NLMS & Channel observations & Systematic exploration of input sizes; ML-based and adaptive \\
        
        Zhao et al.~\cite{zhao2024lstm} & \checkmark & LSTM-based classification & Sub-6 GHz CSI & Reduced input size (4 vs. 16); better training efficiency \\
        Lim et al.~\cite{lim2021deep} & \checkmark & LSTM + Bayesian filtering & Estimated channel states & Lower computational complexity; avoids sequential filtering \\
        
        Alwakeel et al.~\cite{alwakeel20256g} & \checkmark & Virtualized ML-based beamforming & Historical SNR data & We focus on beam tracking and generalization; their method is for static virtualization in \ac{LOS} only \\
        Zhong et al.~\cite{zhong2024image} & \checkmark & CNN for image-based coding & BS-perspective images capturing the UE's position & Model-agnostic to specific spatial layouts; generalizable \\
        Suzuki et al.~\cite{suzuki2022ray} & \checkmark & CNN & LIDAR & Lighter input and more adaptable to general scenarios\\
        Oliveira et al.~\cite{oliveira2024tracking} & \checkmark & CNN + Multimodal fusion & LIDAR, GNSS, beam history & Requires less data, supports general scenarios \\
        Mollah et al.~\cite{mollah2026transformer} & \checkmark & Transformer + multi-modal cross-attention & GPS, camera, LiDAR (DeepSense 6G) & Lighter input (beam RSRP only); no external sensors required; includes NLOS evaluation and measurement overhead analysis \\
        {Jiang et al.~\cite{Alkhateeb_baseline}} & \checkmark & GRU  & LIDAR data and beam historical data(DeepSense 6G) & Lighter input when compared with LiDAR data, and it also takes into account the measurement overhead impact\\
        
        \hline
    \end{tabular}
\end{table*}

\section{Beam Tracking: Problem Formulation, Metrics, and Datasets}
\label{problem_statement}

This section formally describes the problem, detailing the channel model and the beam tracking formulation.

\subsection{Channel Model and Beam Tracking}
\label{sec:beam-tracking}

We assume a \textit{narrowband} channel model, meaning the channel's frequency response is approximately constant over the bandwidth of interest. To capture spatial characteristics, we adopt a geometric-based channel representation incorporating $L$ \ac{MPC}, each associated with specific \ac{AoA} and \ac{AoD}, as well as a complex path gain $\alpha_\ell$~\cite{Heath16}.

The narrowband \ac{MIMO} channel matrix is defined as:
\begin{align}
\mathbf{H} = \sqrt{N_{tx} N_{rx}}\sum_{\ell = 1}^L \alpha_{\ell} \mathbf{a}_r(\phi_\ell^A, \theta_\ell^A)\mathbf{a}^*_t(\phi_\ell^D, \theta_\ell^D),
\label{eq:geometric}
\end{align}
where $N_{tx}$ and $N_{rx}$ denote the number of transmit and receive antenna elements, respectively, $\alpha_\ell$ is the complex gain of the $\ell$-th path, and $\mathbf{a}_t$ and $\mathbf{a}_r$ are the array response vectors for the transmit and receive antennas, which are parameterized by azimuth and elevation angles of departure and arrival, respectively.

\begin{figure}[t]
    \centering
    \includegraphics[width=0.7\columnwidth,trim={0 0.5cm 0 0},clip]{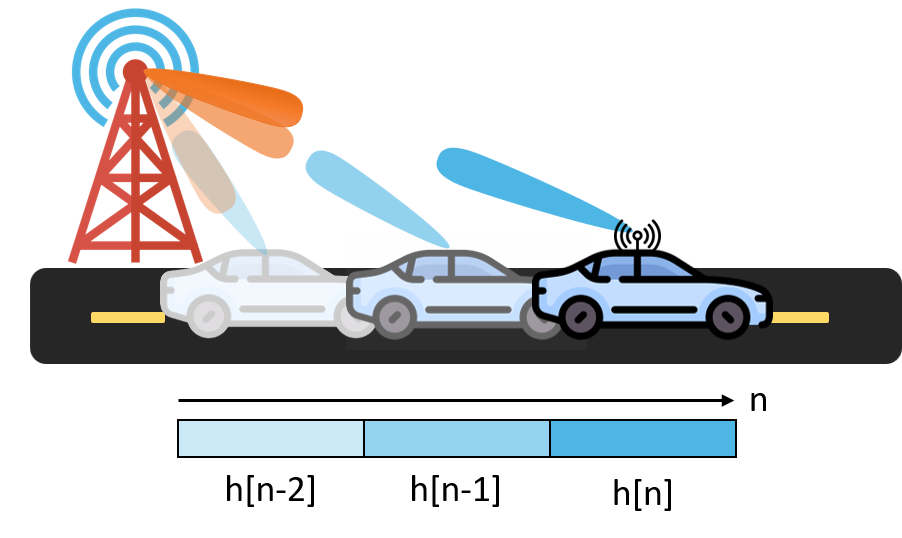}
    \caption{Representation of beam tracking in a vehicular network along the channel $h[n]$ evolution over discrete-time $n$.}
    \label{fig:tracking_example}
    \vspace{-0.3cm}
\end{figure}

This paper addresses beam tracking, \textit{i.e.} the process of dynamically selecting the optimal beam pair in real-time, as transmitter and receiver positions—and consequently, channel conditions—evolve~\cite{giordani2018tutorial}. This is especially crucial in mobile environments where \ac{LOS} paths can become obstructed, and dominant signal paths change rapidly, as illustrated in Figure~\ref{fig:tracking_example}. In this context, efficient beam tracking techniques aim to reduce beam search and alignment overhead, maintain signal quality by aligning with the strongest path, and operate in real time using historical beam sequences or side information.

To analyze and design such techniques, we assume that the beamforming process uses predefined transmit and receive \ac{DFT} codebooks~\cite{he2017codebook}, $\mathcal{C}_T$ and $\mathcal{C}_R$, respectively. It is also assumed that the number of codewords coincides with the number of antenna elements. And for simplicity, given pair $(r,t)$ of beam indices is uniquely identified by a single index $i$, which varies from 1 to $M$, where $M$ is the total number of pairs. Hence, for a given pair of indices, the \emph{combined channel} is
\begin{align}    \label{eq:combined_channel}
    y_i = \mathbf{w}_r^{*} \mathbf{H} \mathbf{f}_t,
\end{align}
where $\mathbf{w}_r$ is the $r$-th combining vector in $\mathcal{C}_R$ and $\mathbf{f}_t$ is the $t$-th precoding vector in $\mathcal{C}_T$.

The core problem is then to identify the beam index
\begin{align}
\hat{i} = \arg\max_{i \in \{1, ..., M\}} |y_i|
\end{align}
that maximizes the magnitude $|y_i|$ of the combined channel. Continuously searching over all $M$ possible beam pairs via sweeping incurs high latency and computational cost, motivating the development of learning-based beam tracking models that predict the optimal beam from sequential and contextual features to reduce overhead and adapt to dynamic environments.

\subsection{Evaluation Metrics}
\label{subsec:metrics}
To assess the performance of the proposed beam tracking algorithms, we employ a set of complementary metrics that capture different aspects of model behavior and its applicability in realistic scenarios. In this study, we consider: a) \ac{RSRP}, b) \ac{TR}, and c) Top-\textit{K} Accuracy—each offering complementary insights into model behavior and its real-world applicability.

\textbf{a) Reference Signal Received Power.}
\ac{RSRP} has been proposed as a key metric for beam tracking in \ac{3GPP} discussions on AI/ML-based beam management for next-generation networks~\cite{3gpp2407554}. \ac{RSRP} measures the received signal strength of the reference signal at the physical layer, providing an indication of the signal quality for a given beam. In beam tracking, \ac{RSRP} is commonly employed to evaluate the performance of different beams and select the optimal one that maximizes the signal power received at the \ac{UE}.

The \ac{RSRP} for a given beam $i$ can be expressed as:
\begin{equation}
\text{RSRP}_i = \frac{1}{N_\mathrm{RS}} \sum_{n=1}^{N_\mathrm{RS}} P_{\mathrm{RS},i}^{(n)},
\end{equation}
where $N_\mathrm{RS}$ is the number of reference signal (RS) resources, and $P_{\mathrm{RS},i}^{(n)}$ is the power of the $n$-th reference signal for beam $i$, resulting in a value proportional to the 
combined channel, as defined in Equation~(\ref{eq:combined_channel}). By periodically computing \ac{RSRP} across multiple beam pairs, the system can identify the beam with the highest signal strength and dynamically switch to it, ensuring a stable and efficient communication link, even in scenarios involving user mobility or varying environments.

\textbf{b) Throughput Ratio.}
The \ac{TR} is a metric used to evaluate the efficiency of beam-tracking algorithms in selecting beam indices for data transmission. It quantifies the ratio between the achievable throughput when using the predicted beam and the maximum possible throughput obtained by selecting the best beam. Mathematically, it is defined as:
\begin{equation}
\textrm{TR} = \frac{\sum_{i=1}^{N_{ts}} \log_2 (1 + y_{\widehat{(i)}})}{\sum_{i=1}^{N_{ts}} \log_2 (1 + y_{\hat i})},
\label{eq:rate_ratio}
\end{equation}
where $N_{ts}$ is the number of test examples, $\hat i$ is the best beam index, and $\widehat{(i)}$ is the predicted beam index.

\textbf{c) Top-\textit{K} Accuracy}. It is widely used to evaluate model performance in multi-class classification and ranking tasks, particularly when considering prediction uncertainty or when exact top-1 accuracy is too restrictive. See \cite{bengio2015deep} for a detailed definition of this metric.

Each metric offers a distinct lens through which the performance of the beam tracking solution can be assessed. RSRP serves as a core indicator of signal quality and is particularly relevant for regression-based evaluations. Tracking Ratio (TR) captures system-level effects by measuring the continuity of successful tracking, offering insight into real-world viability under constrained measurement conditions. Top-\textit{K} Accuracy, while traditionally used in classification tasks, is adapted here to assess regression-based predictions by ranking beam candidates according to their estimated gain; it quantifies how often the optimal beam index appears among the top-\textit{K} predicted candidates.

By integrating these three metrics, we provide a holistic evaluation framework that captures signal fidelity, prediction robustness, and end-user impact—ensuring that the beam tracking solution is assessed not only in terms of model accuracy but also in the context of practical system performance.

\subsection{Dataset Overview}
\label{sec:dataset_description}

This work employs three datasets—\textbf{M10\%}, \textbf{R10\%}, and \textbf{R50\%}—generated using high-fidelity ray-tracing simulations in distinct urban environments. The letter prefix indicates the scenario: \textbf{R} for Rosslyn and \textbf{M} for Marseille. The accompanying percentage represents the proportion of \ac{NLOS} links in the dataset, reflecting different propagation challenges.

\textbf{M10\%}: Represents a low-density, residential-like setting in Marseille, characterized by smooth mobility patterns and predominantly \ac{LOS} propagation. This scenario serves as a baseline due to its relatively stable and predictable dynamics.

\textbf{R10\%}: Models a dense urban canyon in Rosslyn with limited \ac{NLOS} conditions but highly dynamic mobility. The scenario features frequent direction changes and fast link transitions, posing a challenge for tracking algorithms despite the low NLOS ratio.

\textbf{R50\%}: Captures a more balanced urban environment in Rosslyn, with a 50\% mix of \ac{LOS}/\ac{NLOS} links and intermediate mobility dynamics.

\begin{figure}[t]
    \centering
    \includegraphics[width=0.8\columnwidth]{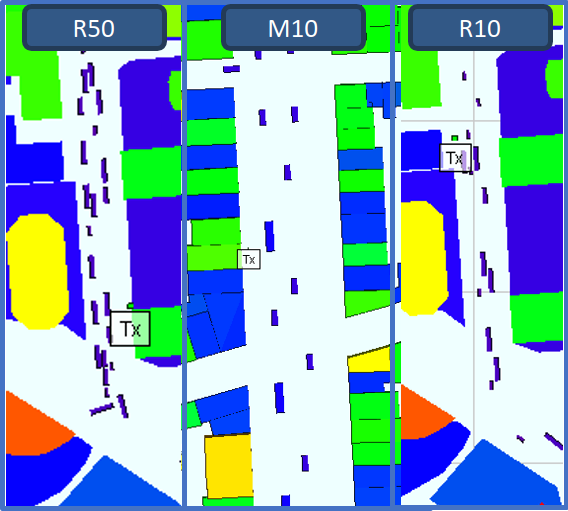}
    \caption{Aerial visualization of the propagation environments in Wireless Insite.}
    \label{fig:scenarios figure}
    \vspace{-0.3cm}
\end{figure}

Figure \ref{fig:scenarios figure} presents an aerial view of the scenarios for each dataset. As stated before, both R10\% and R50\% are situated in Rosslyn, but their \ac{BS} positions differ. These datasets were selected to span a wide range of vehicular communication conditions, enabling thorough stress testing of the proposed beam tracking system in both favorable and adverse environments.

\textbf{a) Line-of-Sight and Non-Line-of-Sight Proportions:}
The LOS/NLOS distribution characterizes the level of signal obstruction present in each dataset. While \ac{LOS} conditions typically allow easier beam alignment, \ac{NLOS} scenarios introduce multi-path propagation and diffraction challenges. Table~\ref{table:Los-Nlos} summarizes the sample distribution for each dataset.

As shown, both M10\% and R10\% datasets contain a high proportion of \ac{LOS} scenes, consistent with 3GPP deployment assumptions~\cite{3gpp2306199} regarding \ac{BS} antenna height (25 meters). In contrast, the \textbf{R50\%} dataset was deliberately configured with a lower antenna height of 10 meters, increasing the likelihood of obstructions and resulting in a significantly higher proportion of \ac{NLOS} links. This setup provides a more challenging environment for beam tracking, making R50\% particularly suitable for assessing algorithm robustness under dynamic and obstructed conditions.

\textbf{b) Beam Dynamics Metrics:}
To analyze how beam preferences evolve over time, we examined multiple complementary metrics. However, due to space constraints, we report only the \ac{MAFD} here.

\textbf{Mean Absolute First Difference}. \ac{MAFD} captures average angular movement between successive beam indices.
Let \( R \) be the number of receivers, \( S \) the number of scenes, and \( N_{\text{total}} \) the total number of beams, the \ac{MAFD} is then computed as:
\begin{equation}
\text{MAFD} = \frac{1}{R} \sum_{r=1}^{R} \frac{1}{S-1} \sum_{s=1}^{S-1} \mathbf{D}_{s_r},
\end{equation}
with
\begin{equation}
\mathbf{D}_{s_r}=\min( (\hat i_{s_r - 1} - \hat i_{s_r}) \text{\%} N_\text{total}, (\hat i_{s_r}-\hat i_{s_r - 1}) \text{\%} N_\text{total}).
\end{equation}
The \%$N_{total}$ operation implements a ``wrap around'', with $N_{total}$ being the total number of beams. 
For instance, assuming $N_{total}=3$, a given receiver $r=5$, and episodes with $S=7$ scenes and a sequence of beam indices $\hat i$ given by $1, 1, 0, 2, 1, 2, 0$;
the respective sequence $\mathbf{D}_{s_5}$ is $0, 1, 1, 1, 1, 1$.

Note the wrap around $N_{total}-1=2$ in this case, such that the last value of  $\mathbf{D}_{s_5}$ is 1 (not 2).

\begin{table}[t]
    \centering
    \caption{LOS/NLOS distribution across datasets}
    \label{table:Los-Nlos}
    \begin{tabular}{|c|c|c|}
        \hline
        \textbf{Dataset} & \textbf{LOS Samples} & \textbf{NLOS Samples} \\
        \hline
        R50\% (Rosslyn) & 14,982 & 13,306 \\
        M10\% (Marseille) & 27,800 & 2,200 \\
        R10\% (Rosslyn) & 27,899 & 2,101 \\
        \hline
    \end{tabular}
\end{table}

The computed values for \ac{MAFD} 
are \textbf{2.04 }for R50\% (Rosslyn), \textbf{0.92} for M10\% (Marseille) and \textbf{1.97} for R10\% (Rosslyn).
Since the dataset is structured hierarchically, composed of multiple episodes each containing several scenes, these metric values represent the average behavior calculated across all episodes. MAFD provides insight into how abruptly and frequently the optimal beam direction changes, which directly impacts tracking difficulty.

\textbf{c) Key Observations and Dataset Justification:}
The observed (MAFD) values reveal clear distinctions in beam dynamics across the evaluated scenarios:
1) \textbf{M10\% (Marseille, 10\% NLOS)} exhibits the lowest MAFD of 0.92, indicating smooth beam dynamics and relatively stable propagation conditions. This setting serves as a baseline for evaluating model performance under ideal or near-ideal tracking scenarios. 2) \textbf{R10\% (Rosslyn, 10\% NLOS)} has a moderately higher MAFD of 1.97, suggesting greater angular variation. The increased complexity likely stems from urban structural features, despite having the same NLOS ratio as M10\%. 3) \textbf{R50\% (Rosslyn, 50\% NLOS)} shows the highest MAFD at 2.04, reflecting substantial angular fluctuations due to dense obstructions and frequent LOS blockages. This makes it the most challenging environment for beam tracking, stressing model robustness under dynamic urban conditions.

MAFD differences underscore the importance of dataset diversity for evaluating tracking models under varied mobility and propagation conditions, enabling realistic and rigorous assessment of model's generalization from stable to complex and rapidly changing environments.

\textbf{d) Data Partitioning:}
To prevent temporal information leakage, the datasets are partitioned at the \emph{episode level} rather than at the individual sample level. Each episode corresponds to a continuous vehicular trajectory, and entire episodes are assigned exclusively to training, validation, or test sets. Specifically, 70\% of episodes are allocated to training, 15\% to validation, and 15\% to testing. This ensures that temporally correlated samples from the same trajectory never appear in both training and evaluation sets, preserving the integrity of the performance assessment. Within each episode, the sliding window described in Section~\ref{subsec:setB} is applied to construct sequential input samples.

\textbf{Reproducibility Note:} To support reproducibility and enable further experimentation, all datasets used in this study are publicly available through our research group website. \footnote{Datasets are available at  \url{https://www.lasse.ufpa.br/pt/raymobtime}, published under the following identifiers:  M10\% (Marseille) — originally \texttt{t005}, R50\% (Rosslyn) — originally \texttt{t004} and R10\% (Rosslyn) — originally \texttt{t006}}
\section{Methodology and Architecture}
\label{sec:proposed_architecture}

This section introduces the proposed beam-tracking framework and outlines the experimental setup designed to evaluate its performance. The system is developed to operate under dynamic vehicular conditions and focuses on two key prediction tasks: beam index classification and \ac{RSRP} regression. The architecture is benchmarked against existing approaches, including the LSTM-based model by Zhao et al.\cite{zhao2024lstm}, a RNN-based approach using historical beam data in a \ac{GRU} model proposed by  Jiang et al.\cite{Alkhateeb_baseline}, and a LIDAR-based CNN approach originally proposed in\cite{dias2019position}, enhanced in~\cite{targ2016resnet, salehi2022deep}, and finalized by Suzuki et al.\cite{suzuki2022ray}.

The novelty of the proposed framework lies not in a single architectural component, but in the integration of several design choices tailored to the beam tracking problem: (i)~a compact input representation using only beam-domain \ac{RSRP} values and beam indices, avoiding high-dimensional raw \ac{CSI} or sensor data; (ii)~a dual-task formulation that enables both classification-based and regression-based beam tracking within the same architecture, exposing distinct trade-offs between accuracy and deployability; and (iii)~an autoregressive inference strategy that leverages the regression model's continuous outputs to substitute real measurements, achieving significant overhead reduction---a capability not available to classification-only approaches. Together, these elements form a cohesive system designed for practical deployment under resource constraints.

The deep learning framework proposed in this work predicts the optimal beam index in real time using a sequence of historical beam measurements. It integrates spatial filtering with temporal modeling via a \ac{RNN} composed of \ac{LSTM} units, allowing the system to adapt to the temporal dynamics of vehicular communication scenarios.

\subsection{Time-Series Input Construction and Overhead Reduction}
\label{subsec:setB}

\begin{figure}[t] \centering \includegraphics[width=0.8\columnwidth,clip]{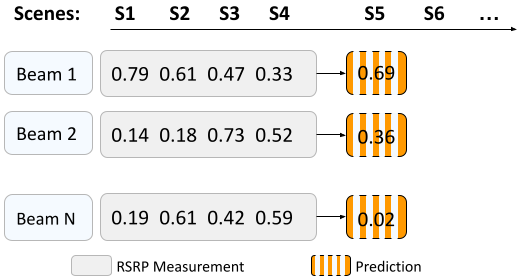} \caption{Structure of historical beam data for time-series models.} \label{fig:arch_pred}
\vspace{-0.3cm}
\end{figure}

Historical measurements and predictions are collected to construct the time-series input for the model. As illustrated in Figure~\ref{fig:arch_pred}, the system employs a sliding window. The framework is flexible, and future research is encouraged to explore different window lengths to better adapt to varying mobility patterns or application constraints. Each time step may contain either a measured or a predicted RSRP value.

\begin{figure}[b] \centering \includegraphics[width=0.8\columnwidth,clip]{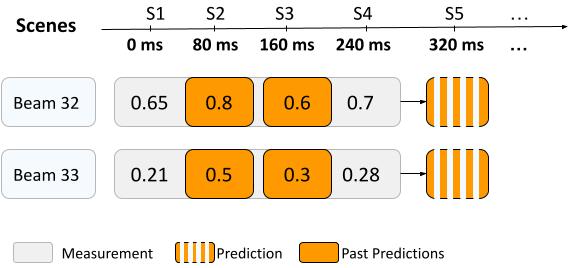} \caption{Example input using real and predicted RSRP values for beams 32 and 33.} \label{fig:SetB_example} \end{figure}

To minimize sensing overhead without compromising tracking quality, the system alternates between actual measurements (every 240 ms) and predictions (every 80 ms), preserving temporal continuity. As shown in Figure~\ref{fig:SetB_example}, this results in mixed sequences of measured and predicted \ac{RSRP} values, as exemplified for beams 32 and 33.

\begin{figure}[ht]
\includegraphics[width=\columnwidth,clip]{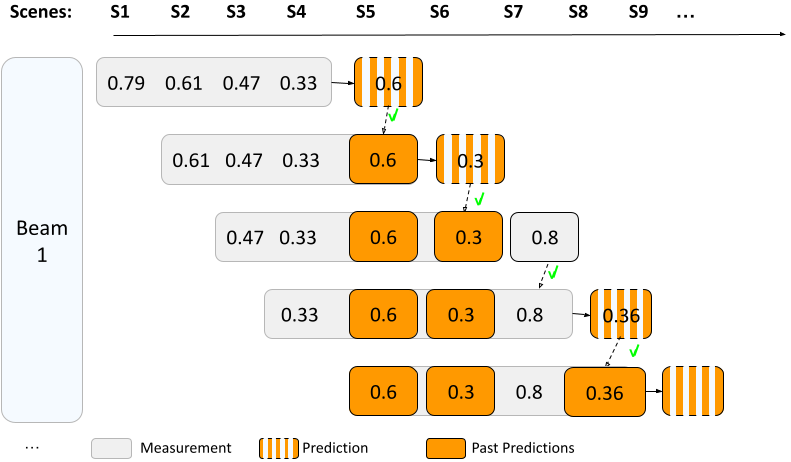}
\caption{Prediction-aided strategy for reducing measurement overhead.}
\label{fig:fig_replacing_Arch}
\vspace{-0.3cm}
\end{figure}

By relying on predictions to fill gaps between measurements, the approach greatly reduces the overhead compared to beam sweep or frequent beam measurements in the worst-case scenarios. For example, if beam measurements were performed at the same frequency as the predictions (i.e., every 80 ms), the overhead would triple. In contrast, the proposed framework, depicted in Figure~\ref{fig:fig_replacing_Arch}, combining one real measurement with two predictions over a 240 ms window—reduces the measurement overhead by approximately 66.7\%, depending on the number of beams being tracked and the channel coherence.

This approach balances prediction freshness and resource efficiency to maintain performance in dynamic scenarios.

\subsection{Prediction Architecture and Tasks}
\label{subsec:arch}
\label{subsec:tasks}


The predictive model leverages a \ac{RNN} architecture based on \ac{LSTM} units, which are well-suited for modeling temporal dependencies inherent in beam tracking behavior. LSTMs overcome the vanishing gradient limitations of conventional RNNs through gated mechanisms—specifically, input, forget, and output gates—that regulate the flow of information across time steps. This structure enables the model to retain relevant context over extended sequences, making it particularly effective for beam tracking in mobile communication systems, where signal quality evolves continuously over time.

Formally, let $\mathbf{x}[n] \in \mathbb{R}^{M}$ denote the vector of \ac{RSRP} measurements (or predictions) across all $M$ candidate beams at discrete time step $n$, where each entry $x_i[n]$ corresponds to the received power $|y_i[n]|^2$ for beam pair $i$ as defined in Eq.~\eqref{eq:combined_channel}. The model receives as input a temporal sequence $\mathbf{X} = [\mathbf{x}[n-W+1], \ldots, \mathbf{x}[n]] \in \mathbb{R}^{W \times M}$, where $W$ is the observation window length.

For the classification task (DeepBT-C), the model learns a mapping $f_C\colon \mathbb{R}^{W \times M} \rightarrow \Delta^{M-1}$, where $\Delta^{M-1}$ is the $(M{-}1)$-simplex, producing a probability distribution over beam indices. The predicted beam is then:
\begin{equation}
\hat{i}_{n+1} = \arg\max_{i \in \{1,\ldots,M\}} \; [f_C(\mathbf{X})]_i.
\label{eq:classification_pred}
\end{equation}

For the regression task (DeepBT-R), the model learns a mapping $f_R\colon \mathbb{R}^{W \times M} \rightarrow \mathbb{R}^{M}$, estimating the \ac{RSRP} for each beam at the next time step. The predicted beam is selected as:
\begin{equation}
\hat{i}_{n+1} = \arg\max_{i \in \{1,\ldots,M\}} \; [f_R(\mathbf{X})]_i.
\label{eq:regression_pred}
\end{equation}

In the autoregressive inference mode (Section~\ref{sec:autoregressive_inference}), the predicted output $\hat{\mathbf{x}}[n+1] = f_R(\mathbf{X})$ is fed back into the input sequence, replacing the corresponding real measurement. This yields the updated input $\mathbf{X}' = [\mathbf{x}[n-W+2], \ldots, \mathbf{x}[n], \hat{\mathbf{x}}[n+1]]$ for the subsequent prediction step.

\begin{figure}[b]
\centering
\includegraphics[width=0.8\columnwidth,trim={0 1cm 0 1cm},clip]{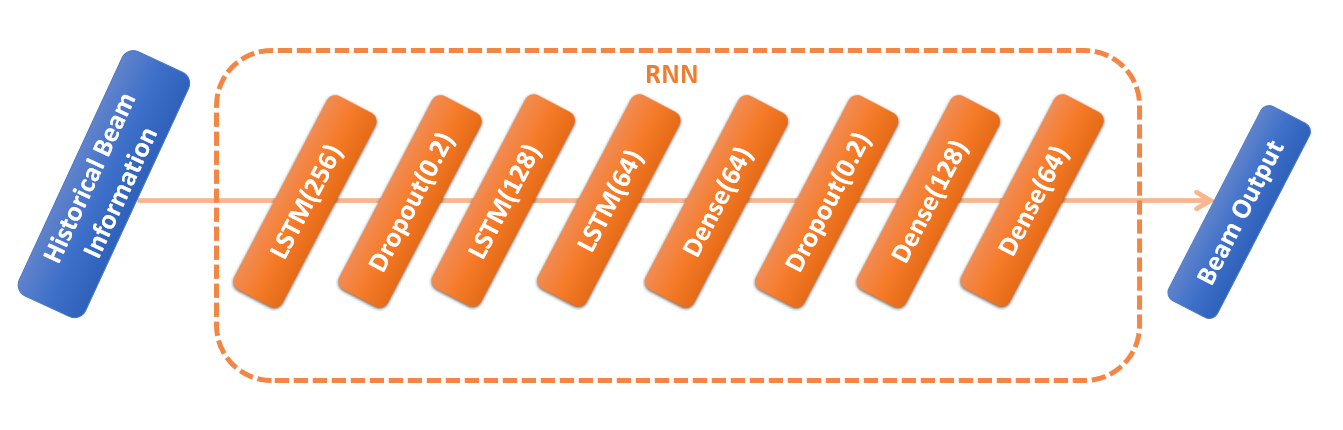} \caption{Proposed beam tracking architecture based on LSTM layers.} \label{fig:Architecture} \end{figure}

The architecture of the model, shown in Figure~\ref{fig:Architecture}, consists of an input layer that encodes the time-series beam data; LSTM layers that extract sequential features from historical RSRP values and beam indices; dropout layers for regularization to reduce overfitting; and dense layers that map the extracted features to the final output predictions.

The architecture supports two core learning tasks:

\textbf{1. Deep Learning Beam Tracking via Classification -- DeepBT-C}
The model predicts the most likely beam index for the next time step based on current time-series inputs. The output is a probability distribution over beams, and metric Top-K Accuracy is used to evaluate performance.

\textbf{2. Deep Learning Beam Tracking via Regression -- DeepBT-R}
The model estimates the RSRP for each beam. The beam with the highest predicted RSRP is selected as optimal. The model is trained with Mean Squared Error (MSE) loss. To unify the evaluation with classification-based models, Top-K accuracy is also applied to regression predictions by identifying in each prediction the beam index with the highest predicted gain and checking whether this index falls within the \textit{K} highest-gain beams in the ground-truth vector. This approach preserves the core intuition of Top-\textit{K} Accuracy, evaluating whether the model's top-ranked prediction aligns with the most favorable beams in terms of actual signal strength.

\subsection{Baseline Methods}

To provide a comparative analysis, the proposed deep learning model was evaluated against three baseline methods relevant to \ac{mmWave} beam management and tracking:

\textbf{1. Zhao et al. LSTM}: This deep learning-based approach targets beam tracking in co-located scenarios using \ac{LSTM} layers trained on historical sequences of sub-6 GHz CSI~\cite{zhao2024lstm}. Each time step in the input sequence contains the complex channel coefficients between the \ac{UE} and the \ac{BS}. Our evaluation employs a re-implementation of this model, adapted to the datasets used in this study. It is important to note that reported performance in the original work or other studies may differ due to variations in datasets, input pre-processing, or evaluation conditions.

\textbf{2. Jiang et al. GRU}: This model leverages gated recurrent units (GRUs) to perform beam tracking using a real-world dataset (DeepSense 6G)~\cite{Alkhateeb_baseline}. Their approach achieves up to 97.6\% Top-5 accuracy under \ac{LOS} conditions and serves as a strong benchmark. However, the evaluation assumes perfect knowledge of previous beam selections and does not consider \ac{NLOS} scenarios, which limits its applicability in more complex or obstructed environments. We reproduce this method on our datasets to ensure consistent comparison under equivalent assumptions.

\textbf{3. Suzuki et al. CNN}: This baseline is a \ac{CNN}-based approach that leverages spatial sensing data, such as LIDAR, for beam selection and tracking~\cite{suzuki2022ray}. We adopted this method as our sensing-based baseline instead of the LIDAR extension proposed by Jiang et al.~\cite{Alkhateeb_baseline}, primarily due to its compatibility with the Raymobtime dataset used in this study.

\section{Experimental Results}
\label{sec:experimental_results}

This section presents a comprehensive evaluation of the proposed RNN-based beam tracking framework, assessing both task variants—\textbf{DeepBT-R} and \textbf{DeepBT-C} described in Section~\ref{subsec:tasks}. The evaluation is conducted across three datasets with varying propagation conditions, identified by their \ac{NLOS} percentages: \textbf{M10\%} — Marseille, 10\% \ac{NLOS} (90\% LOS), \textbf{R10\%} — Rosslyn, 10\% NLOS (90\% LOS) and \textbf{R50\%} — Rosslyn, 50\% NLOS (balanced scenario).

The experimental analysis begins by evaluating the input efficiency of our proposed models in comparison to baseline architectures. We then investigate the effects of LOS/NLOS transitions on beam stability and prediction accuracy, followed by a detailed assessment of our measurement overhead reduction strategy. Finally, we analyze throughput efficiency, and the trade-offs between performance and sensing overhead.

\subsection{Model Specifications and Input Efficiency}
Table~\ref{table:model_specifications} presents a comparison of the proposed DeepBT models against three baseline architectures in terms of parameter count (i.e., the total number of trainable weights and biases), model size, and average input size. Although the DeepBT model has a comparable number of parameters to the CNN baseline by Suzuki et al.~\cite{suzuki2022ray}, it operates on input vectors that are over 800 times smaller than Suzuki et al.'s approach, highlighting its significant advantage in terms of memory efficiency and real-time feasibility. 

When comparing the proposed model to the LSTM and GRU-based baselines by Zhao et al.\cite{zhao2024lstm} and Jiang et al.\cite{Alkhateeb_baseline}, respectively, we observe a different trade-off. Although these models are smaller in terms of parameter count and model size, they rely on larger input vectors—potentially increasing CSI acquisition and transmission overhead. This is particularly relevant in uplink scenarios, where channel state information is gathered at the UE and must be transmitted to the BS. In contrast, the DeepBT model leverages minimal input while preserving competitive model capacity, making it more efficient in both computation and communication contexts.

Although the proposed model is not the smallest in terms of parameter count, its compact input size makes it an attractive candidate for real-time, resource-constrained beam tracking. Future work may explore architecture optimizations to reduce model complexity without compromising accuracy and its input efficiency advantage.

\begin{table}[ht]
    \caption{Model Specifications}
    \label{table:model_specifications}
    \centering
    \begin{tabular}{|c|c|c|c|}
        \hline
        \textbf{Model} & \textbf{Parameters} & \textbf{Model Size} & \textbf{Mean Input Size} \\
        \hline
        DeepBT & 535,552 & 2.04 MB & 1.25 KB \\
        Zhao et al. LSTM & 55,872 & 0.218 MB & 5 KB \\
        Jiang et al. GRU & 43,008 & 0.168 MB & 1.6 KB \\
        Suzuki et al. CNN & 514,000 & 1.96 MB & 1048 KB \\
        \hline
    \end{tabular}
\end{table}

\subsection{Comparison Against Baseline Models}

\begin{figure}[t]
    \centering
    \includegraphics[width=0.8\columnwidth]{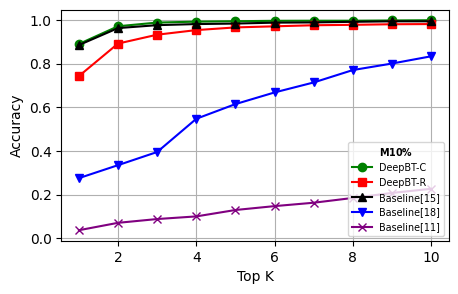}
    \includegraphics[width=0.8\columnwidth]{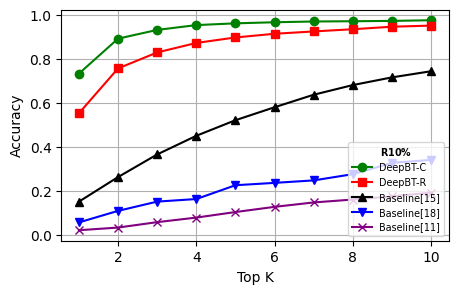}
    \caption{Top-$K$ accuracy comparison within proposed models vs. baselines in datasets M10\% and R10\% (10\% NLOS)}
    \label{fig:los_baselines}
    \vspace{-0.3cm}
\end{figure}

Figure~\ref{fig:los_baselines} presents a comparison of Top-$K$ accuracy between our proposed models—DeepBT-C (Classification) and DeepBT-R (Regression)—and several baselines across two datasets: M10\% and R10\%, which consist of 90\% LOS and 10\% NLOS scenes. In both scenarios, DeepBT-C consistently achieves the highest accuracy across all $K$ values, outperforming classical and recent LSTM-based baselines~\cite{zhao2024lstm}. DeepBT-R also performs competitively, slightly behind DeepBT-C but still significantly ahead of all baselines, particularly in the more challenging R10\% dataset.

While DeepBT-R underperforms relative to DeepBT-C in terms of raw accuracy, it enables the application of an autoregressive inference strategy~\cite{rubio2010structural}, explained in Section~\ref{subsec:arch}, that can substantially reduce beam tracking overhead during deployment. This makes DeepBT-R particularly attractive for real-time, resource-constrained scenarios. The benefits of this technique are further discussed in Section~\ref{sec:autoregressive_inference}.

Notably, while the LIDAR baseline~\cite{suzuki2022ray} shows reasonable performance in R10\%, they remain far below the accuracy achieved by our deep learning-based approaches. These results highlight the robustness and generalization capabilities of our method under LOS-dominant conditions and suggest that even minor NLOS occurrences do not significantly degrade performance. In the following section, we evaluate model performance in more complex environments, specifically the R50\% dataset with a balanced LOS/NLOS distribution.

\subsection{Impact of LOS/NLOS Transitions on Beam Stability and Model Performance}
\label{sec:NLOS_impact}

\begin{figure}[b]
\centering
\includegraphics[width=\columnwidth]{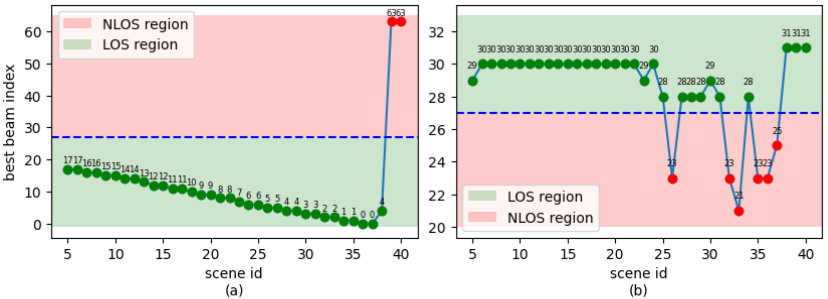}
\caption{Temporal evolution of the optimal beam index across LOS and NLOS intervals in R10\% and R50\%.}
\label{fig:nlos_examples}
\end{figure}

To assess the influence of propagation conditions on beam tracking, we analyze model behavior under mixed LOS/NLOS dynamics using the R50\% dataset, which includes 50\% NLOS scenes. As illustrated in Figure~\ref{fig:nlos_examples}, abrupt beam index changes frequently occur in NLOS intervals due to blockages and multi-path propagation. These transitions contrast sharply with the more stable index behavior observed during LOS, where beam direction remains largely consistent.

Figure~\ref{fig:nlos_examples} shows two representative temporal patterns. In subplot (a), the beam index changes gradually, with one abrupt switch near the end due to an NLOS event. Subplot (b)
depicts a more erratic behavior: long intervals of stable beams are interrupted by brief but severe disruptions caused by transient LOS blockages. These examples show how short NLOS episodes can pose significant prediction challenges.

\begin{figure}[b]
\centering
\includegraphics[width=0.8\columnwidth]{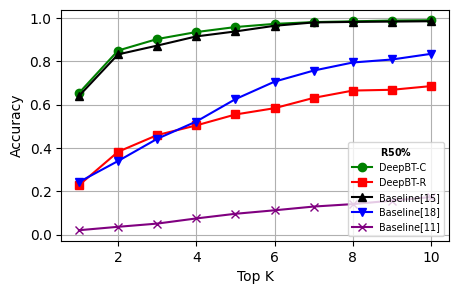}
\caption{Top-$K$ accuracy comparison in R50\% (50\% NLOS): proposed models vs. baselines.}
\label{fig:nlos_baselines}
\end{figure}

To evaluate model robustness under frequent propagation changes, we compare performance on the challenging R50\% dataset, which features a balanced 50\% LOS and 50\% NLOS distribution. As shown in Figure~\ref{fig:nlos_baselines}, our proposed models significantly outperform all baselines across all Top-$K$ values, demonstrating strong generalization under mixed conditions.

In this scenario, the \textbf{DeepBT-C} model consistently achieves the highest accuracy, outperforming both DeepBT-R and all baseline methods. DeepBT-R, while slightly behind DeepBT-C, still maintains a clear margin over all baselines. The gap between DeepBT-R and classification grows under more severe NLOS conditions, which highlights the limitations of regression-based approaches in environments with abrupt signal fluctuations.

Notably, the LIDAR-based baseline~\cite{suzuki2022ray} also performs very well in this setting, demonstrating that spatial context can substantially enhance prediction accuracy. However, this comes at the cost of significantly higher input dimensionality and computational overhead, making it less practical for lightweight or latency-sensitive deployments.

These results reinforce the importance of designing models that can adapt to unpredictable real-world conditions. Despite the higher complexity of the R50\% environment, both DeepBT variants maintain high accuracy, with DeepBT-C nearing 99\% Top-10 accuracy and DeepBT-R reaching over 70\%. These findings validate the robustness of our approach for deployment in urban and obstructed scenarios.

\begin{figure}[t]
\centering
\includegraphics[width=0.8\columnwidth]{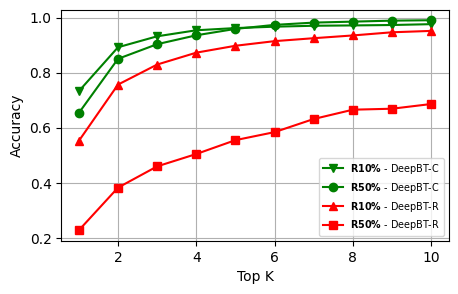}
\caption{Top-$K$ accuracy comparison between RSRP Regression and Best Index Classification in R10\% vs. R50\%.}
\label{fig:r10_vs_r50}
\vspace{-0.3cm}
\end{figure}

Figure~\ref{fig:r10_vs_r50} compares the Top-$K$ accuracy of DeepBT-C and DeepBT-R across the same environment in two different propagation conditions: R10\%  and R50\%. As expected, both models achieve higher accuracy in the more favorable R10\% scenario, where LOS conditions dominate and beam direction changes are smoother. DeepBT-C shows strong resilience, with only a minor drop in performance between the two settings—maintaining over 95\% accuracy for Top-5 predictions even in R50\%. This demonstrates the model’s robustness to propagation variability.

In contrast, DeepBT-R is more sensitive to the increased presence of NLOS in R50\%, exhibiting a noticeable drop in accuracy, particularly for lower $K$ values. Despite this, it still delivers meaningful predictions and benefits from its lower inference overhead, as discussed in Section~\ref{sec:autoregressive_inference}. These results confirm that while classification-based approaches are more reliable under dynamic conditions, regression remains a practical alternative for lightweight inference when paired with techniques like autoregression.

\begin{figure*}[t]
    \centering
    \includegraphics[width=0.85\textwidth]{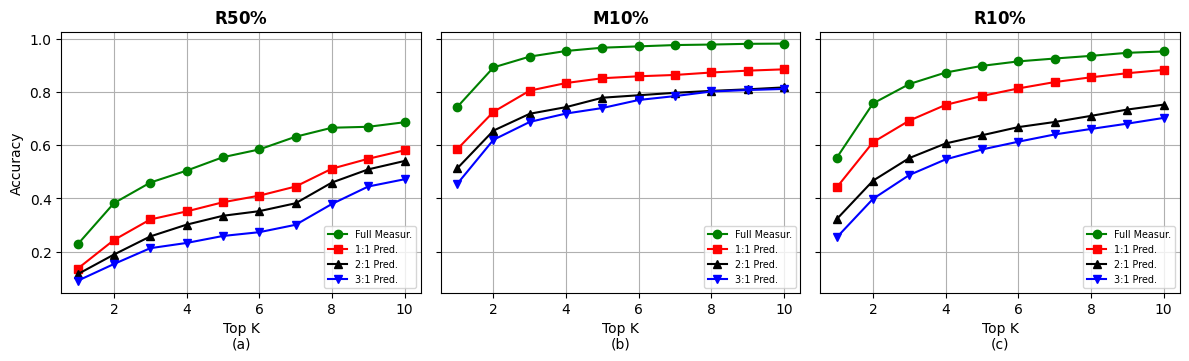}
    \caption{Top-$K$ accuracy under different measurement replacement strategies. First row: Best Index Classification; Second row: RSRP Regression. Top to bottom: (a, b) R50\%, (c, d) M10\%, (e, f) R10\%.}
    \label{fig:fig_replace_class_reg_measur_pred}
    \vspace{-0.3cm}
\end{figure*}

\subsection{Autoregressive Inference for Overhead Reduction}
\label{sec:autoregressive_inference}

This section evaluates the performance of autoregression strategies, where real beam measurements are partially replaced by model predictions in the input sequence. As described in Section~\ref{sec:proposed_architecture} and illustrated in Figure~\ref{fig:fig_replacing_Arch}, this approach allows reducing measurement overhead by reusing recent model outputs in place of actual sensing data.

Specifically, we analyze three replacement configurations: the 1:1 prediction strategy, where each model prediction is followed by one real measurement; the 2:1 prediction strategy, where two consecutive predictions are made before a new measurement is taken; and the 3:1 prediction strategy, which uses three predictions for every one measurement.

Figure~\ref{fig:fig_replace_class_reg_measur_pred} shows the Top-$K$ accuracy of DeepBT-R under these replacement strategies across the three datasets. In Fig.~\ref{fig:fig_replace_class_reg_measur_pred}, we use ``\emph{The Full Measur.}'' to refer to evaluations made only with real measurements, which serve as a baseline. 

As expected, tracking accuracy declines gradually as the number of consecutive predictions increases. However, even in the most challenging scenario (R50\% with 3:1 prediction), DeepBT-R maintains a reasonable level of accuracy. In LOS-dominant datasets (M10\% and R10\%), the impact of replacement is minimal, and the model preserves high accuracy even with fewer measurements. In R10\% scenario, the DeepBT-R even outperforms all baseline methods in the same scenario, achieving results close to the LIDAR baseline.

This trade-off is particularly advantageous for real-time applications such as vehicular beam tracking and mobile edge computing, where sensing time competes with data transmission and control signaling. By reducing the frequency of beam measurements, the system can allocate more resources to communication tasks.

The corresponding \ac{MOR} for each configuration can be quantified following 3GPP Technical Report~\cite{3gpp38843} and prior work~\cite{jayaweera20245g}, using the formula:
\begin{equation}
\text{MOR (\%)} = \left(1 - \frac{N_\text{measured}}{N_\text{full}}\right) \times 100
\end{equation}

\noindent where \(N_\text{measured}\) is the number of actual beam measurements used in a reduced strategy, and \(N_\text{full}\) is the number used in the full-measurement setup. For example, the 1:1, 2:1, and 3:1 configurations correspond to approximately 50\%, 66.7\%, and 75\% reduction in sensing overhead, respectively. 

Figure~\ref{fig:fig_replace_class_reg_measur_pred} shows the impact of this strategy on both proposed models—RSRP Regression and Best Index Classification—across the three datasets. These results confirm that DeepBT-R supports low-overhead autoregressive inference while maintaining strong performance, offering a practical path for efficient beam tracking in deployment-constrained environments.

\textbf{Error Accumulation.}
A known limitation of autoregressive inference is the potential accumulation of prediction errors over successive steps, as each predicted value feeds into the input for the next prediction. This effect is visible in Figure~\ref{fig:fig_replace_class_reg_measur_pred}, where accuracy progressively decreases from the 1:1 to 3:1 configurations. The degradation is more pronounced in the R50\% dataset, where abrupt \ac{NLOS} transitions amplify prediction uncertainty. To mitigate this, the proposed framework periodically resets the input sequence with real measurements (every 240~ms in the 2:1 configuration), effectively bounding the maximum length of consecutive predictions. In practice, selecting the replacement ratio involves a trade-off between overhead reduction and acceptable accuracy loss, and may be adapted dynamically based on channel conditions---an avenue left for future investigation.

\subsection{Throughput Analysis}
\label{subsec:throughput}
\begin{figure}[!t]
    \centering
    \includegraphics[width=0.8\columnwidth]{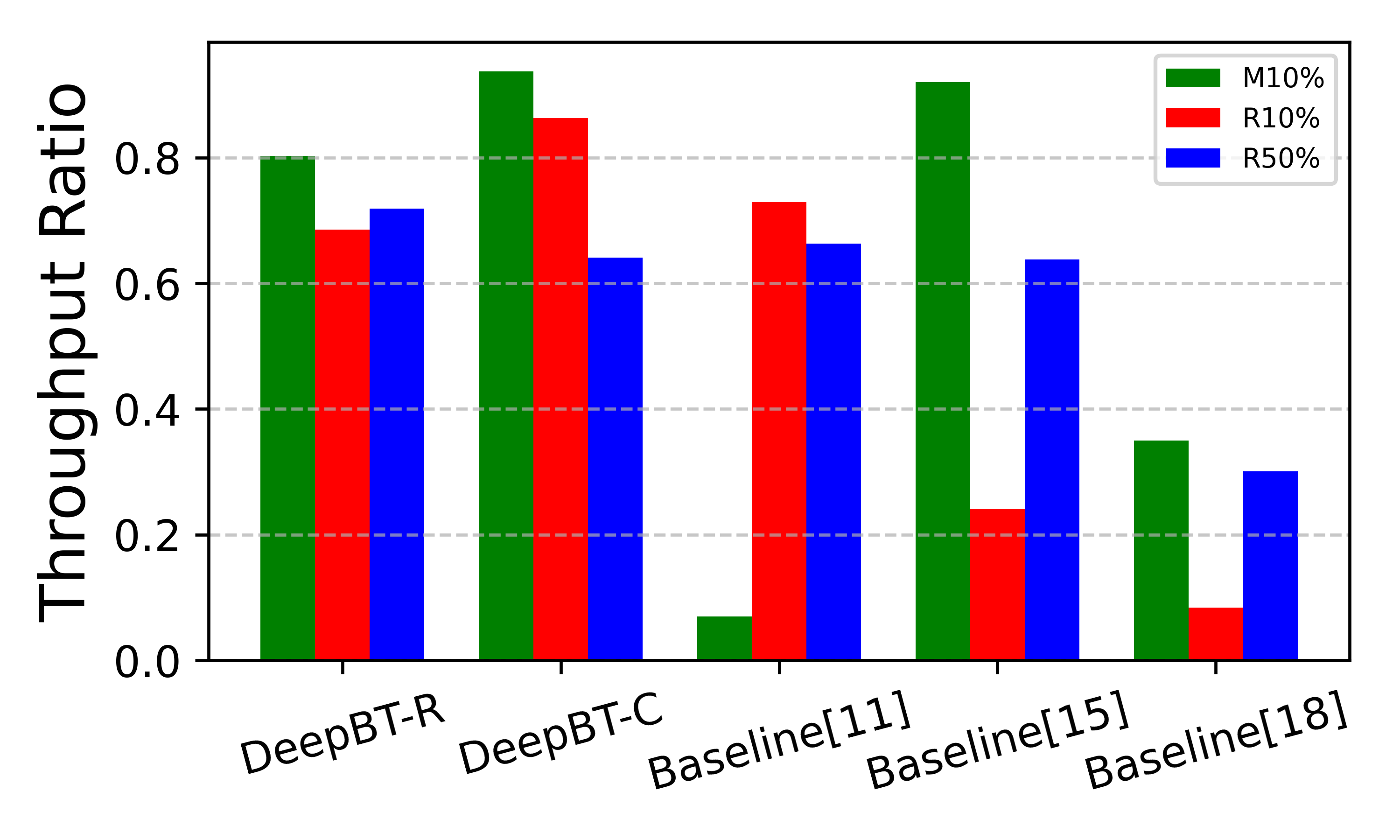}
    \caption{Top-1 Throughput Ratio for RSRP Regression and Best Index Classification models across datasets.}
    \label{fig:t_Ratio}
    \vspace{-0.3cm}
\end{figure}

Figure~\ref{fig:t_Ratio} presents the \ac{TR} for Top-1 accuracy across all evaluated datasets. This metric captures the fraction of successful beam alignments relative to the theoretical maximum, thus serving as a proxy for end-to-end communication efficiency.

Among all models, DeepBT-C consistently achieves high throughput across datasets, particularly under the M10\% and R10\% conditions. DeepBT-R also demonstrates strong performance, especially in R50\%, where it outperforms all baselines by a significant margin, and even surpasses the DeepBT-C in this specific dataset.

Traditional baselines exhibit larger performance gaps, particularly under the M10\% and R10\% scenarios, where their throughput ratios remain substantially lower. This underscores the advantage of the adaptability of DeepBT strategies.

Interestingly, R50\% yields more consistent throughput across all models. This behavior aligns with the lower beam gain variance in this dataset (8.42 dB), which makes beam selection errors less critical, providing more leeway for beam mis-selection without severe performance penalties. In contrast, M10\% and R10\% have larger beam gain variance (14.24 dB and 11.60 dB, respectively), meaning the performance gap between the optimal and suboptimal beams is more pronounced. As a result, accurate beam selection becomes more important, and misclassifications tend to cause sharper drops in throughput.

\section{Conclusion and Future Work}
\label{sec:conclusion}

This work presented a robust deep learning framework for beam tracking in \ac{mmWave} \ac{V2I} scenarios. We conducted extensive evaluations across three realistic datasets with varying LOS/NLOS distributions. The proposed architecture, DeepBT, leverages historical channel dynamics through sequential modeling to support both beam classification and regression strategies. Among them, the DeepBT-R model, based on RSRP regression, achieves a favorable trade-off between prediction accuracy and measurement overhead, while the DeepBT-C outperforms all baselines in accuracy in the analyzed datasets.

Future research will explore more advanced sequential modeling architectures, such as Transformer-based and hybrid RNN-Transformer approaches, to enhance
temporal dependency extraction under fast mobility. Additionally, integrating external sensing modalities (e.g., radar or GNSS) may further improve beam pre-selection, especially in cluttered urban environments. On the efficiency side, promising directions include knowledge distillation, structured pruning of \ac{LSTM} units, quantization-aware training for edge deployment, or replacing the \ac{LSTM} backbone with lighter alternatives such as temporal convolutional networks (TCNs). For completeness, upcoming evaluations will also include non--deep learning benchmark methods. Another important open question is the impact of observation window length $W$ and its interaction with mobility speed and channel coherence time. While a fixed window of four time steps was adopted in this work based on preliminary tuning, exploring adaptive window sizing could
further improve the balance between tracking responsiveness and prediction stability.

An interesting direction for future work involves the integration of position-aware spatial beam pre-selection to reduce search space and computational overhead. Preliminary experiments (now omitted for brevity) explored a filtering strategy in which only beams within a small angular vicinity of the estimated user direction are considered. This approach, grounded in 3GPP beam management guidelines and based on CSI-derived user positioning, showed promising results in reducing inference time with minimal performance degradation, particularly in LOS-dominant environments. A more systematic exploration of adaptive beam set sizing and its impact across diverse propagation scenarios remains a valuable avenue for future research.

\bibliographystyle{IEEEtran}
\bibliography{main}

@article{jayaweera20245g,
  title={{5G}-{A}dvanced {AI/ML} Beam Management: Performance Evaluation with Integrated {ML} Models},
  author={Jayaweera, Nalin and Bonfante, Andrea and Schamberger, Mark and Tehrani, Amir Mehdi Ahmadian and Sanguanpuak, Tachporn and Tilak, Preetish and Jayasinghe, Keeth and Vook, Frederick W and Rajatheva, Nandana},
  journal={arXiv preprint arXiv:2404.15326},
  year={2024}
}

@article{mollah2026transformer,
  author  = {Mollah, Muhammad Baqer and Wang, Honggang and Karim,
             Mohammad Ataul and Fang, Hua},
  title   = {Multi-Modal Sensing and Fusion in mmWave Beamforming for
             Connected Vehicles: A Transformer Based Framework},
  journal = {IEEE Transactions on Vehicular Technology},
  year    = {2026},
  publisher = {IEEE}
}

@techreport{3gpp38843,
  organization = {{3GPP}},
  title        = {Study on Artificial Intelligence ({AI})/Machine Learning ({ML}) for {NR} Air Interface},
  institution  = {3rd Generation Partnership Project (3GPP)},
  type         = {Technical Report},
  number       = {TR 38.843},
  year         = {2023},
  month        = {December},
  note         = {Version 18.0.0},
  url          = {https://www.3gpp.org/ftp/Specs/archive/38_series/38.843/38843-1800.zip}
}

@book{bengio2015deep,
  title={Deep Learning},
  author={Bengio, Yoshua and Goodfellow, Ian and Courville, Aaron},
  year={2015},
  publisher={MIT Press}
}

@techreport{beam_Sweeping_3gpp.38.802,
  title = {{Study on New Radio Access Technology: Physical Layer Aspects}},
  institution = {3rd Generation Partnership Project (3GPP)},
  number = {TR 38.802},
  year = {2017},
  note = {Version 14.2.0}
}

@techreport{beam_management_csi_3gpp.38.213,
  title = {{NR; Physical layer procedures for control}},
  institution = {3rd Generation Partnership Project (3GPP)},
  number = {TS 38.213},
  year = {2024},
  note = {Version 18.3.0}
}

@article{alwakeel20256g,
  title={6G virtualized beamforming: a novel framework for optimizing massive MIMO in 6G networks},
  author={Alwakeel, Ahmed M},
  journal={EURASIP Journal on Wireless Communications and Networking},
  volume={2025},
  number={1},
  pages={23},
  year={2025},
  publisher={Springer}
}

@article{salehi2022deep,
  title={{Deep learning on multimodal sensor data at the wireless edge for vehicular network}},
  author={Salehi, Batool and Reus-Muns, Guillem and Roy, Debashri and Wang, Zifeng and Jian, Tong and Dy, Jennifer and Ioannidis, Stratis and Chowdhury, Kaushik},
  journal={IEEE Transactions on Vehicular Technology},
  volume={71},
  number={7},
  pages={7639--7655},
  year={2022},
  publisher={IEEE}
}

@article{rubio2010structural,
  title={Structural vector autoregressions: Theory of identification and algorithms for inference},
  author={Rubio-Ramirez, Juan F and Waggoner, Daniel F and Zha, Tao},
  journal={The Review of Economic Studies},
  volume={77},
  number={2},
  pages={665--696},
  year={2010},
  publisher={Wiley-Blackwell}
}

@article{he2017codebook,
  title={{Codebook-based hybrid precoding for millimeter wave multiuser systems}},
  author={He, Shiwen and Wang, Jiaheng and Huang, Yongming and Ottersten, Bj{\"o}rn and Hong, Wei},
  journal={IEEE Transactions on Signal Processing},
  volume={65},
  number={20},
  pages={5289--5304},
  year={2017},
  publisher={IEEE}
}

@ARTICLE{Heath16,
  author={R. W. {Heath} and N. {González-Prelcic} and S. {Rangan} and W. {Roh} and A. M. {Sayeed}},
  journal={IEEE Journal of Selected Topics in Signal Processing}, 
  title={{An Overview of Signal Processing Techniques for Millimeter Wave MIMO Systems}}, 
  year={2016},
  volume={10},
  number={3},
  pages={436-453},
}

@misc{3gpp2306199,
  author       = {{3GPP}},
  title        = {{Feature lead summary \#3 evaluation of AI/ML for beam management}},
  howpublished = {Meeting Document R1-2306199, 3GPP TSG RAN WG1 Meeting \#112b-e, e-Meeting},
  month        = apr,
  year         = {2023},
}

@misc{3gpp2407554,
  author       = {{3GPP}},
  title        = {{FL summary \#5 for AI/ML in beam management}},
  howpublished = {Meeting Document R1-2407554, 3GPP TSG RAN WG1 Meeting \#118, Maastricht, NL},
  month        = apr,
  year         = {2023},
}

@article{giordani2018tutorial,
  title={{A tutorial on beam management for {3GPP NR} at {mmWave} frequencies}},
  author={Giordani, Marco and Polese, Michele and Roy, Arnab and Castor, Douglas and Zorzi, Michele},
  journal={IEEE Communications Surveys \& Tutorials},
  volume={21},
  number={1},
  pages={173--196},
  year={2018},
  publisher={IEEE}
}

@article{targ2016resnet,
  title={{Resnet in resnet: Generalizing residual architectures}},
  author={Targ, Sasha and Almeida, Diogo and Lyman, Kevin},
  journal={arXiv preprint arXiv:1603.08029},
  year={2016}
}

@inproceedings{suzuki2022ray,
  title={{Ray-Tracing {MIMO} Channel Dataset for Machine Learning Applied to {V2V} Communication}},
  author={Suzuki, Daniel and Oliveira, Ailton and Gon{\c{c}}alves, Luan and Correa, Ilan and Klautau, Aldebaro and Lins, Silvia and Batista, Pedro},
  booktitle={2022 IEEE Latin-American Conference on Communications (LATINCOM)},
  pages={1--6},
  year={2022},
  organization={IEEE}
}

@article{Alkhateeb_baseline,
  title={LiDAR aided future beam prediction in real-world millimeter wave {V2I} communications},
  author={Jiang, Shuaifeng and Charan, Gouranga and Alkhateeb, Ahmed},
  journal={IEEE Wireless Communications Letters},
  volume={12},
  number={2},
  pages={212--216},
  year={2022},
  publisher={IEEE}
}

@inproceedings{dias2019position,
  title={{Position and {LIDAR}-aided {mmWave} beam selection using deep learning}},
  author={Dias, Marcus and Klautau, Aldebaro and Gonz{\'a}lez-Prelcic, Nuria and Heath, Robert W},
  booktitle={2019 IEEE 20th International Workshop on Signal Processing Advances in Wireless Communications (SPAWC)},
  pages={1--5},
  year={2019},
  organization={IEEE}
}

@article{zhao2024lstm,
  title={{LSTM-Based Predictive mmWave Beam Tracking via Sub-6 GHz Channels for {V2I} Communications}},
  author={Zhao, Yao and Zhang, Xianchao and Gao, Xiaozheng and Yang, Kai and Xiong, Zehui and Han, Zhu},
  journal={IEEE Transactions on Communications},
  year={2024},
  publisher={IEEE}
}

@article{xue2024survey,
  title={{A survey of beam management for mmWave and THz communications towards 6G}},
  author={Xue, Qing and Ji, Chengwang and Ma, Shaodan and Guo, Jiajia and Xu, Yongjun and Chen, Qianbin and Zhang, Wei},
  journal={IEEE Communications Surveys \& Tutorials},
  year={2024},
  publisher={IEEE}
}

@article{Bjornson19,
title={{Massive MIMO in Sub-6 GHz and mmWave: Physical, Practical, and Use-Case Differences}},
author={E. {Bjornson} and L. {Van der Perre} and S. {Buzzi} and E. G. {Larsson}},
journal={IEEE Wireless Commun.},
year={2019},
volume={26},
number={2},
pages={100-108}
}

@article{Roh14,
title={{Millimeter-wave beamforming as an enabling technology for 5G cellular communications: theoretical feasibility and prototype results}},
author={W. {Roh} and J. {Seol} and J. {Park} and B. {Lee} and J. {Lee} and Y. {Kim} and J. {Cho} and K. {Cheun} and F. {Aryanfar}},
journal={IEEE Communications Magazine},
year={2014},
volume={52},
pages={106-113}
}

@article{mmWave2020,
  title={{Multi-Frequency Multi-Scenario Millimeter Wave MIMO Channel Measurements and Modeling for B5G Wireless Communication Systems}}, 
  author={Huang, Jie and Wang, Cheng-Xiang and Chang, Hengtai and Sun, Jian and Gao, Xiqi},
  journal={IEEE Journal on Selected Areas in Communications}, 
  year={2020},
  volume={38},
  pages={2010-2025}
}

@article{yi2024beam,
  title={{Beam training and tracking in mmwave communication: A survey}},
  author={Yi, Wang and Zhiqing, Wei and Zhiyong, Feng},
  journal={China Communications},
  year={2024},
  publisher={IEEE}
}

@article{zhong2024image,
  title={{Image-Based Beam Tracking With Deep Learning for mmWave V2I Communication Systems}},
  author={Zhong, Weizhi and Zhang, Lulu and Jin, Haowen and Liu, Xin and Zhu, Qiuming and He, Yi and Ali, Farman and Lin, Zhipeng and Mao, Kai and Durrani, Tariq S},
  journal={IEEE Transactions on Intelligent Transportation Systems},
  year={2024},
  publisher={IEEE}
}

@article{lim2021deep,
  title={{Deep learning-based beam tracking for millimeter-wave communications under mobility}},
  author={Lim, Sun Hong and Kim, Sunwoo and Shim, Byonghyo and Choi, Jun Won},
  journal={IEEE Trans. Commun.},
  volume={69},
  number={11},
  pages={7458--7469},
  year={2021},
  publisher={IEEE}
}

@inproceedings{shaham2020extended,
  title={{Extended Kalman filter beam tracking for millimeter wave vehicular communications}},
  author={Shaham, Sina and Kokshoorn, Matthew and Ding, Ming and Lin, Zihuai and Shirvanimoghaddam, Mahyar},
  booktitle={2020 IEEE International Conference on Communications Workshops (ICC Workshops)},
  pages={1--6},
  year={2020},
  organization={IEEE}
}

@article{asi2021beam,
  title={{Beam tracking channel for millimeter-wave communication system using least mean square algorithm}},
  author={Asi, Ban A and Mohmood, Farhad E},
  journal={Al-Rafidain Engineering Journal (AREJ)},
  volume={26},
  number={2},
  pages={118--123},
  year={2021},
  publisher={Al-Rafidain Engineering Journal (AREJ)}
}

@INPROCEEDINGS{oliveira2024tracking,
  author={Oliveira, Ailton and Suzuki, Daniel and Bastos, Sávio and Correa, Ilan and Klautau, Aldebaro},
  booktitle={2024 IEEE Latin-American Conference on Communications (LATINCOM)}, 
  title={Machine Learning-Based mmWave MIMO Beam Tracking in {V2I} Scenarios: Algorithms and Datasets}, 
  year={2024},
  volume={},
  number={},
  pages={1-5},
  doi={10.1109/LATINCOM62985.2024.10770674}}

\end{document}